\documentclass{easychair}

\usepackage{doc}
\usepackage{amsmath}
\usepackage{amssymb}
\usepackage{amsthm}
\usepackage{enumitem}

\newcommand{\easychair}{\textsf{easychair}}

\title{A Privacy-Preserving Information-Sharing Protocol for Federated Authentication}

\author{
Francesco Buccafurri\thanks{Corresponding author}
\and 
Carmen Licciardi
}

\institute{
University Mediterranea of Reggio Calabria,
  Reggio Calabria, Italy\\
  \email{bucca,carmen.licciardi@unirc.it}
 }

\authorrunning{Buccafurri et al.}

\titlerunning{The {\easychair} Class File}

\begin{document}

\maketitle

\begin{abstract}
This paper presents a privacy-preserving protocol for identity registration and information sharing in federated authentication systems. The goal is to enable Identity Providers (IdPs) to detect duplicate or fraudulent identity enrollments without revealing users’ personal data or enabling cross-domain correlation. The protocol relies on Oblivious Pseudorandom Functions (OPRFs) combined with domain-specific transformations, ensuring that each IdP generates independent pseudonymous identifiers derived from a shared cryptographic service while maintaining full input confidentiality. A central authority maintains a blind registry that records successful and failed identity verifications using only pseudonymous identifiers, allowing global consistency checks without exposing sensitive information or linking users across domains. The proposed construction provides a general and abstract framework suitable for a wide range of federated authentication systems, achieving strong privacy guarantees while supporting effective fraud-prevention mechanisms during identity registration.
\end{abstract}

\section{Introduction}

Federated authentication has become a fundamental paradigm for enabling seamless and secure access to distributed digital services. In this model, multiple independent Identity Providers (IdPs) authenticate users and issue assertions or tokens that Service Providers (SPs) can rely on without performing additional verification. This arrangement improves usability, reduces the proliferation of credentials, and allows organizations to maintain control over the authentication of their users while supporting interoperability across administrative domains.

Despite these advantages, federated authentication systems face structural challenges during the identity registration phase. Each IdP performs user onboarding autonomously, often relying on local procedures and heterogeneous verification levels. This decentralization creates opportunities for attackers to exploit inconsistencies across IdPs, attempting multiple registrations under the same real-world identity or impersonating legitimate users. At the same time, enabling IdPs to exchange detailed identity attributes to detect such behavior would violate core privacy principles and regulatory requirements aimed at minimizing personal data exposure.

Identity verification typically relies on Know Your Customer (KYC) procedures, where an IdP validates the user’s identity through official documents, biometric evidence, or secure digital credentials. While KYC ensures accurate verification within each individual domain, it does not inherently prevent an attacker from registering the same identity at multiple IdPs. Without a coordination mechanism, repeated or fraudulent onboarding attempts remain undetected across the federation.

These limitations highlight a fundamental tension: preventing duplicate or fraudulent identity registrations requires some form of information sharing among IdPs, yet such sharing must not reveal personal data or enable cross-domain user tracking. As a result, designing privacy-preserving mechanisms for identity registration and information exchange has become an essential research direction in modern federated systems.

In this paper, we present a protocol that addresses this gap by enabling IdPs to coordinate identity registration through a central service while preserving user privacy. The key idea is to allow IdPs to submit blinded, domain-specific representations of user identifiers that cannot be linked or reversed by any party. A central coordinating service maintains a registry of these pseudonymous identifiers and can therefore detect whether an identity has been previously registered or whether a past verification attempt has failed, all without learning anything about the underlying personal data or correlating users across different IdPs.
The solution preserves the autonomy of IdPs, respects the privacy constraints of modern regulatory frameworks, and introduces a novel, privacy-preserving mechanism for global identity consistency checks during registration.

The structure of the paper is the following. 
Section~\ref{sec:background} introduces the relevant background on federated authentication and discusses the challenges that motivate our work. 
Section~\ref{sec:model} presents the system model, the involved entities, and the notation used throughout the paper.  
Section~\ref{sec:protocol} describes the proposed protocol, including first-time identity enrollment, subsequent registrations, and the cooperative global check mechanism.  
Section~\ref{sec:related} reviews related work, and 
Section~\ref{sec:conclusion} concludes the paper and outlines directions for future research.

\section{Background}
\label{sec:background}

\subsection{Oblivious Pseudorandom Functions}

An \textit{Oblivious Pseudorandom Function (OPRF)} enables a client to obtain $F_K(x)$ from a server holding the secret key $K$, such that:
\begin{itemize}[noitemsep]
    \item The client learns only $F_K(x)$, not $K$;
    \item The server learns nothing about $x$.
\end{itemize}

The OPRF guarantees that from the perspective of the client, $F_K(x)$ is computationally indistinguishable from the output of a random function, provided the server is honest.

\subsection{RSA-Based Instantiation}

Let the server hold an RSA key pair $K = (N, e, d)$.
The OPRF can be instantiated as:
\[
F_K(x) = H'(x^d \bmod N)
\]
where $H'$ is a hash mapping RSA outputs into a fixed domain.  
To compute $F_K(x)$ obliviously, the client performs:
\[
\begin{aligned}
X &= H(x)\cdot r^e \bmod N, \\
Y &= X^d \bmod N, \\
F_K(x) &= Y \cdot r^{-1} \bmod N.
\end{aligned}
\]
This ensures that $F_K(x) = (H(x))^d \bmod N$ without revealing $x$ to the server.

\subsection{Blind Signatures}

Blind signatures~\cite{chaum1982blind} extend RSA to sign messages without revealing them to the signer.
Given $m$, the requester blinds it as $m' = m \cdot r^e \bmod N$, the signer returns $(m')^d$, and the requester unblinds to obtain $m^d$.
This ensures unlinkability between signing and verification.

\section{Notation and Entities}
\label{sec:model}

The protocol involves three entities:
\begin{itemize}[noitemsep]
    \item \textbf{User}: owns personal data including a unique personal identifier ($\text{UPI}$) 
    and undergoes KYC at an IdP. For cryptographic purposes, the user is represented 
    by the hashed value $x = H(\text{UPI})$, which is revealed only to the IdP performing 
    the registration.
    
    \item \textbf{Identity Provider (IdP$_i$)}: verifies KYC, blinds $x$, and applies a 
    domain-specific RSA transformation using its private exponent $t_i$. Each IdP$_i$ 
    holds an RSA key pair $(N_i, e_i, t_i)$ defining its domain, while the CTS stores 
    only the public component $(N_i, e_i)$.

    \item \textbf{Central Trusted Service (CTS)}: holds the master RSA key pair 
    $K = (N, e, d)$, evaluates blinded inputs, issues blind-signed tokens, and maintains 
    a pseudonymous registry storing minimal KYC status information.
\end{itemize}

Whenever $x = H(\text{UPI})$ must be sent to the CTS, the IdP blinds it using a random 
factor $r \in \mathbb{Z}_N^*$, ensuring that no party other than the originating IdP 
learns the value of $H(\text{UPI})$. All OPRF evaluations by the CTS operate modulo $N$, so 
every pseudonymous identifier $PID_i$ produced during enrollment also lies in 
$\mathbb{Z}_N$.

IdP$_i$ uses its private exponent $t_i$ to derive a domain-specific pseudonymous 
identifier $PID_i$ through a cooperative RSA-OPRF interaction with the CTS. The same 
exponent is employed during the cooperative check phase, where IdP$_i$ computes 
$r^{t_i} \bmod N_i$ to enable unblinding of identifiers belonging to its domain.

Following a successful KYC verification at IdP$_i$, the CTS issues a blind signature 
over $(PID_i, pk_u)$, generating a token $Token_i$ that binds the pseudonymous 
identifier to the user’s public key. During subsequent registrations, the user proves 
ownership of the corresponding private key via a challenge–response procedure.

The blind registry maintained by the CTS stores only a minimal verification flag 
$status_{KYC} \in \{ok,\, alarm\}$ for each pseudonym. No additional data is stored or 
revealed, ensuring strict privacy and data minimization.

The complete set of symbols used in the protocol is summarized in 
Figure~\ref{fig:notations}.

\begin{figure}[t!]
\centering
\begin{tabular}{ll}
\hline
\textbf{Symbol} & \textbf{Meaning} \\ 
\hline

$\text{UPI}$ & User’s unique personal identifier. \\[4pt]

$x = H(\text{UPI})$ & Hash of the identifier used for cryptographic processing. \\[4pt]

$r$ & Blinding factor in $\mathbb{Z}_N^*$. \\[4pt]

$t_i,\; (N_i,e_i)$ & RSA private exponent $t_i$ and public key of IdP $i$. \\[4pt]

$K=(N,e,d)$ & RSA key pair of the CTS. \\[4pt]

$Blind(x)=x\cdot r^e \bmod N$ & RSA blinding function (CTS modulus). \\[4pt]

$Eval_K(X)=X^d \bmod N$ & OPRF evaluation performed by the CTS. \\[4pt]

$PID_i = (H(\text{UPI}))^{\,d \cdot t_i} \bmod N$ 
& Pseudonymous identifier for IdP $i$ (always modulo $N$). \\[4pt]

$Token_i = \text{Sign}_{CTS}(PID_i, pk_u)$ 
& Blind signature binding user public key $pk_u$ to $PID_i$. \\[4pt]

$status_{KYC} \in \{ok,\, alarm\}$ & Status stored in the blind registry. \\[2pt]

\hline
\end{tabular}
\caption{Notation used throughout the protocol.}
\label{fig:notations}
\end{figure}

\section{Protocol Definition}\label{sec:protocol}
In this section, we present the privacy-preserving protocol that enables identity
registration and consistency checks across a federated authentication system.
The protocol is articulated into three main phases, each corresponding to a
different state in which a user may interact with an Identity Provider (IdP).

In the first phase, a user with no prior credentials undergoes an initial enrollment,
during which a domain-specific pseudonymous identifier is generated and an initial
token is issued. This token serves as proof that the user has successfully completed
a verified onboarding process at least once within the federation.

In the second phase, users performing subsequent registrations present their existing
token to demonstrate prior successful verification. The IdP then generates a new
pseudonymous identifier for its own domain while preserving unlinkability with past
registrations.

In cases where no valid token is available, the IdP must determine—without learning
or disclosing any sensitive information—whether the same user has already been
registered at another provider. To accomplish this, the protocol incorporates a
cooperative blind global check that allows IdPs to query a central coordinating
service in a privacy-preserving manner.

The following subsections describe these three phases in detail.

\subsection{First-Time Identity Enrollment}

Before a user can obtain a registered identity for the first time, they must undergo an
initial enrollment procedure at an Identity Provider (IdP).  
This phase is special because the user does not yet possess any cryptographic token
issued by the CTS, and therefore cannot prove prior verification.  
Consequently, the IdP must rely solely on the outcome of its own KYC process and on 
the cooperative RSA-OPRF interaction with the CTS to generate the user’s first 
pseudonymous identifier.

The objective of this phase is twofold:  
(i) to bind the user’s real-world identity (validated through KYC) to a domain-specific 
pseudonymous identifier $PID_1$, and  
(ii) to let the CTS issue the user’s first blind-signed token.  
Importantly, the token is issued only after the user proves possession of the private
key corresponding to the public key $pk_u$ included in the message to be signed.
This ensures that the token cannot be transferred or misused by an attacker.

The steps of the first-time enrollment are as follows.

\begin{enumerate}[noitemsep]
    \item The user provides $\text{UPI}$ to IdP$_1$ and successfully completes the KYC procedure.
    \item IdP$_1$ computes $x = H(\text{UPI})$, selects a random blinding factor $r$, 
          and computes the blinded input $Blind(x)=x\cdot r^e \bmod N$.
    \item The IdP applies its domain transformation by computing 
          $X_1 = Blind(x)^{t_1} \bmod N$ and sends $X_1$ to the CTS.
    \item The CTS evaluates the OPRF by computing 
          $Y_1 = X_1^d \bmod N$ and returns it to IdP$_1$.
    \item IdP$_1$ unblinds:
          \[
          PID_1 = Y_1 \cdot (r^{t_1})^{-1} \bmod N.
          \]
    \item The user (via IdP) sends $(PID_1, pk_u)$ to the CTS, and proves possession of 
        the private key associated with $pk_u$ through a challenge--response protocol.
        Only if this proof succeeds, the CTS stores $(PID_1, \text{status} = \text{ok})$ in its registry
        and issues to the user the blind-signed $Token_1 = \text{Sign}_{CTS}(PID_1, pk_u)$.
\end{enumerate}

\subsection{Subsequent Registration}

After the user has successfully completed at least one identity enrollment, the CTS has
already issued to them a blind-signed token $Token_j$ for some $j$.  
Any future enrollment at a different Identity Provider (IdP) follows a similar
cryptographic workflow to the first-time registration, with two important differences:
(i) the user is expected to prove that they have already been verified at least once, and 
(ii) the IdP must rely on this proof to distinguish between a new legitimate enrollment 
and a potential impersonation attempt.

To register again at an IdP, the user begins by undergoing the same RSA-OPRF procedure
to compute a new domain-specific pseudonymous identifier $PID_i$, exactly as in the
initial enrollment.  
Before the IdP proceeds with KYC, the user must present a previously issued token
$Token_j$ together with a proof of possession of the corresponding private key.  
This is achieved through a standard challenge--response mechanism: the IdP sends
a fresh challenge and the user returns its signature using the private key associated
with the public key contained in $Token_j$.  
Since all valid tokens bind the same user public key $pk_u$ to distinct pseudonyms,
any previously issued token is acceptable, regardless of its age or of which IdP
issued it.

This mechanism allows IdPs to verify that the user attempting a subsequent registration 
has already been successfully enrolled at least once, while preserving unlinkability:
the IdP cannot determine when, where, or how many times the user was previously enrolled,
and the CTS never learns the user’s unique personal identifier or personal information.
Clearly, if the registration to IdP$_i$ succeeds, then a token $Token_i$ is released to the user.

\subsection{Cooperative Blind Global Check Protocol}
\label{sec:coopcheck}

When a user attempts a registration at IdP$_i$ without presenting a valid token, the 
IdP must determine whether the same unique personal identifier has already been successfully 
registered at any other IdP, without ever learning any identifier outside its own 
domain.  
This is achieved through a cooperative blind global check involving IdP$_i$, all other 
IdPs, and the CTS.  
All computations are performed without exposing $H(\text{UPI})$, any PID belonging to 
another IdP, or any sensitive information.

The protocol consists of three conceptual steps, in which multiple entities 
contribute to the computation.

\subsection*{Blinded Submission}

IdP$_i$ begins by computing the blinded value
\[
X = H(\text{UPI})\cdot r^e \bmod N,
\]
where $r \in \mathbb{Z}_N^*$ is random and known only to IdP$_i$.  
The value $X$ hides $H(\text{UPI})$ from the CTS and from all other IdPs.  
IdP$_i$ sends $X$ to the CTS.

\subsection*{Distributed Domain Transformations}

To determine whether the same user has been registered under any IdP$_j$, the CTS 
forwards $X$ to each IdP$_j$.  
Each IdP$_j$ applies its domain exponent and returns:
\[
T_j = X^{t_j} \bmod N.
\]
Since all computations are modulo $N$, these values are compatible with the OPRF 
evaluation performed by the CTS.  
The CTS computes
\[
Y_j = T_j^d \bmod N
\]
for every IdP$_j$ and returns $Y_j$ to IdP$_i$.  
At this point, the CTS has evaluated the blinded input under every IdP domain 
without learning the underlying user identity.

\subsection*{Exponent Exchange}

To locally unblind $Y_j$, IdP$_i$ must obtain the value $r^{t_j}$, but IdP$_j$ should not 
learn anything about $r$ or $H(\text{UPI})$.  
The following RSA-based exchange allows IdP$_i$ to obtain $r^{t_j}$ blindly, using 
IdP$_j$’s own modulus $N_j$ only for this auxiliary step.

For each IdP$_j$:
\begin{enumerate}[noitemsep]
    \item IdP$_i$ samples a random $k \in \mathbb{Z}_{N_j}^*$ and computes
          \[
          m = k^{e_j}\cdot r \bmod N_j.
          \]
    \item IdP$_i$ sends $m$ to IdP$_j$.
    \item IdP$_j$ computes
          \[
          s = m^{t_j} \bmod N_j = k \cdot r^{t_j} \bmod N_j
          \]
          and returns $s$ to IdP$_i$.
    \item IdP$_i$ recovers
          \[
          r^{t_j} = s \cdot k^{-1} \bmod N_j.
          \]
\end{enumerate}

This step reveals no information about $H(\text{UPI})$ or about the PIDs across domains, 
while giving IdP$_i$ exactly the quantity needed for unblinding.

\subsection*{Local Unblinding and Registry Match}

Given $Y_j$ and $r^{t_j}$, IdP$_i$ reconstructs the pseudonymous identifier for the 
domain of IdP$_j$:

\[
PID_j = Y_j \cdot (r^{t_j})^{-1} \bmod N.
\]

IdP$_i$ sends all reconstructed identifiers $\{PID_j\}$ to the CTS.  
The CTS checks whether any of them appears in its registry with status \emph{ok}.  

If such a match exists, this indicates that the attempted registration corresponds 
to an identity already verified at another IdP, and IdP$_i$ must reject the current 
procedure.  
In that case, $(PID_i,\, status = alarm)$ is inserted into the CTS registry.

If no match exists, IdP$_i$ proceeds with enrollment as for a first-time registration.

\section{Related Work}
\label{sec:related}

Research on privacy-preserving identity systems spans several decades, beginning with
the foundational notions of blind signatures~\cite{chaum1982blind} and pseudonymous
transactions~\cite{chaum1985security}. These early primitives introduced the idea of
obtaining attestations without revealing underlying data, paving the way for
cryptographic privacy in identity management. Anonymous credential systems, most
notably the Camenisch--Lysyanskaya construction~\cite{camenisch2001efficient},
further developed these concepts by enabling unlinkable and non-transferable
credentials. While such systems provide strong privacy guarantees during
authentication, they do not address the problem of detecting duplicate or fraudulent
identity registrations across independent domains.

Privacy concerns in federated identity management have also been extensively studied.
Systems such as PRIMA~\cite{asghar2016prima} and related approaches focus on
preventing cross-provider tracking by issuing unlinkable identifiers to users. Broader
analyses of privacy threats in federated systems, including correlation attacks and
metadata leakage, have been discussed in~\cite{danezis2005privacy}. These works,
however, primarily concentrate on the authentication phase and generally assume that
identity enrollment is performed correctly at each provider, without offering
mechanisms for detecting reuse of the same real-world identity across IdPs.

A parallel line of research examines OPRF-based anonymous token systems. Architectures
such as Privacy Pass~\cite{davidson2020privacy} and the emerging VOPRF
standard~\cite{krawczyk2023voprf} demonstrate how OPRFs and blind signatures can be
combined to produce unlinkable tokens suitable for privacy-preserving authorization
and rate-limiting. More advanced schemes, including Anon-Tokens~\cite{silde2022anon}
and OPRF-based adaptive anonymous credentials~\cite{baseri2024adaptive}, illustrate
the practical benefits of integrating OPRFs into credential issuance workflows. These
approaches, however, generally target anonymous access control and do not address
identity proofing or coordinated verification across multiple providers.

Operational and regulatory frameworks such as NIST~SP~800-63-3~\cite{nist80063}
highlight the challenges of identity proofing and the risks associated with inconsistent
enrollment procedures across decentralized identity providers. Current federated
architectures lack mechanisms to detect duplicate or fraudulent registrations without
centralizing sensitive user information, creating a tension between security and
privacy that remains unresolved.

In contrast to prior work, our protocol focuses explicitly on the enrollment phase in
federated identity systems. We introduce a cooperative RSA-OPRF-based mechanism
that enables a central coordinating service to detect duplicate or fraudulent
registrations without ever learning personal identity attributes or enabling
cross-provider correlation. To the best of our knowledge, this is the first approach
to provide global consistency of identity verification outcomes while preserving full
domain separation and strong privacy guarantees across independently operated Identity
Providers.

\section{Conclusion}
\label{sec:conclusion}

We presented a cooperative protocol for privacy-preserving identity registration in
federated authentication systems. The construction enables a central coordinating
service to detect duplicate or fraudulent registrations without learning users’ personal
identifiers or acquiring information that would allow correlation across Identity
Providers (IdPs). By combining blinded evaluations, domain-specific transformations,
and non-transferable verification tokens, the protocol provides a mechanism for
ensuring global consistency of identity verification outcomes while preserving strong
privacy guarantees.

The proposed architecture achieves properties that are rarely simultaneously supported
in existing federated identity systems: unlinkability across IdPs, blindness of the
central authority, and reliable detection of identity reuse, all while preventing the
misuse or transfer of verification tokens. These features contribute to a more robust
foundation for identity enrollment, enabling federated services to coordinate without
exposing sensitive information or relaxing their privacy constraints.

Although the protocol demonstrates that privacy and global consistency can coexist,
several open directions remain. One avenue concerns integrating revocation mechanisms:
legitimate users may need to invalidate compromised tokens, yet revocation must be
performed without creating new correlation vectors. Another direction involves
formalizing the security model under stronger adversarial assumptions, including
malicious IdPs and adaptive attackers; a full cryptographic proof would further
strengthen confidence in the approach. Operational questions also arise: evaluating
scalability in environments with many IdPs and optimizing the multi-domain consistency
check, for example through batching or threshold-based techniques, are important steps
toward real-world deployment.

Finally, aligning this protocol with evolving federated identity architectures—
including emerging digital identity frameworks and wallet-based models—offers a
promising direction for future work. Introducing privacy-preserving coordination at
the enrollment layer may significantly enhance the resilience of federated systems
against identity fraud while maintaining strict protection of user information.

\section*{Acknowledgments}
This work is supported by Agenzia per la Cybersicurezza Nazionale under the programme for promotion of XL cycle PhD research in cybersecurity – C36E24000080005.
The views expressed are those of the authors and do not represent the funding institutions.

\label{sect:bib}
\bibliographystyle{plain}

\end{document}